\documentclass[12pt,preprint]{aastex}

\newcommand{\Rs}{$R_{\rm s}$}
\newcommand{\HR}{H/R}

\newcommand{\al}{$\alpha$}

\newcommand{\omk}{$\Omega_{\rm K}$}
\newcommand{\mdot}{$\dot{M}$}
\newcommand{\Tmax}{$T_{\rm max}$}

\def\sles{\lower2pt\hbox{$\buildrel {\scriptstyle <}
   \over {\scriptstyle\sim}$}}
\def\sgreat{\lower2pt\hbox{$\buildrel {\scriptstyle >}
   \over {\scriptstyle\sim}$}}
\begin{document}

\title{VISCOUS TORQUE AND DISSIPATION IN THE INNER REGIONS OF A THIN
ACCRETION DISK: IMPLICATIONS  FOR MEASURING BLACK HOLE SPIN}

\author{Rebecca Shafee\altaffilmark{1},
Ramesh Narayan\altaffilmark{2},
Jeffrey E. McClintock\altaffilmark{2}}

\altaffiltext{1}{Harvard University, Department of Physics,
17 Oxford  Street, Cambridge, MA 02138}
\altaffiltext{2}{Harvard-Smithsonian Center for Astrophysics, 60 Garden
Street, Cambridge, MA 02138}

\begin{abstract}

We consider a simple Newtonian model of a steady accretion disk around
a black hole.  The model is based on height-integrated hydrodynamic
equations, $\alpha$-viscosity, and a pseudo-Newtonian potential which
results in an innermost stable circular orbit (ISCO) that closely
approximates the one predicted by GR. We find that, as the disk
thickness $H/R$ or the value of $\alpha$ increases, the hydrodynamic
model exhibits increasing deviations from the standard thin disk model
of Shakura \& Sunyaev. The latter is an analytical model in which the
viscous torque is assumed to vanish at the ISCO.  We consider the
implications of the results for attempts to estimate black hole spin
by using the standard disk model to fit continuum spectra of black
hole accretion disks.  We find that the error in the spin estimate is
quite modest so long as $H/R \leq 0.1$ and $\alpha \leq 0.2$.  At
worst the error in the estimated value of the spin parameter is 0.1
for a non-spinning black hole; the error is much less for a rapidly
spinning hole.  We also consider the density and disk thickness
contrast between the gas in the disk and that inside the ISCO. The
contrast needs to be large if black hole spin is to be successfully
estimated by fitting the relativistically-broadened X-ray line profile
of fluorescent iron emission from reflection off an accretion disk.
In our hydrodynamic models, the contrast in density and thickness is
low when $H/R\ \sgreat\ 0.1$, suggesting that the iron line technique
may be most reliable in extremely thin disks.  We caution that these
results have been obtained with a viscous hydrodynamic model. While 
our results are likely to be qualitatively correct, quantitative
estimates of, e.g., the magnitude of the error in the spin estimate,
need to be confirmed with MHD simulations of radiatively cooled thin
disks.

\end{abstract}


\normalsize

\section{INTRODUCTION}

Recently, we reported spin estimates of three black holes (BHs) in
Galactic X-ray binaries (Shafee et al. 2006; McClintock et al. 2006;
hereafter S06, M06). The results were obtained by fitting the soft
X-ray continuum spectra of these systems in the thermal state
(Remillard \& McClintock 2006) to a general relativistic, multicolor
blackbody, thin disk model ($Kerrbb$, Li et al. 2005), which includes
the effect of spectral hardening (Davis et al. 2005). In this method,
which was pioneered by Zhang, Cui \& Chen (1997), we assume a
razor-thin disk that terminates at the innermost stable circular orbit
(ISCO). In addition, we assume that the viscous torque vanishes at the
ISCO and that there is no energy dissipation or angular momentum loss
inside the ISCO.  These are standard assumptions in the theory of
accretion disks (e.g., Shakura \& Sunyaev 1973; Frank, King \& Raine
2002), and correspond to what we refer to in this paper as the
``standard disk model.''  However, there has been debate in recent
times as to the validity of the assumptions.

The stress responsible for angular momentum transport in a thin
accretion disk is likely to be magnetic (Balbus \& Hawley 1991).  If
this is the case, an argument could be made for a non-zero stress at
the ISCO, coupled with considerable dissipation near and inside the
ISCO (Krolik 1999; Gammie 1999).  These effects could cause
important deviations from the standard disk model (Krolik \& Hawley
2002), perhaps invalidating our spin determinations.

Afshordi \& Paczy\'nski (2003), following earlier work by
Abramowicz \& Kato (1989) and Paczy\'nski (2000), suggested that the
torque at the ISCO increases with increasing disk thickness.
Motivated by their work, we argued in M06 that deviations from the
standard disk model are likely to be serious only for thick disks.  We
thus restricted our attention to relatively thin disks with
height-to-radius ratios of $\HR<0.1$.  The present paper is an attempt
to verify whether or not such thin disks do indeed behave like the
standard disk model.

In addition to the debate over the validity of using the standard
disk theory to model the continuum spectra of realistic disks,
another relevant issue in attempting to estimate BH spin is the
relative merit of the continuum fitting method compared to fitting
the relativistically-broadened fluorescent iron line in the X-ray
spectrum. Both methods have been proposed as a means of estimating
BH spins, and it is of interest to understand how well the
assumptions of each are satisfied by real disks.  The models
currently used by the iron line method assume that the line
emissivity peaks at the ISCO, drops abruptly to zero inside the
ISCO, and decreases steeply as a broken power-law outside the ISCO
(e.g., Brenneman \& Reynolds 2006, hereafter BR06). This requires,
among other things, a significant drop in matter density (Fabian
2007) or disk thickness (Nayakshin et al. 2000, 2002) inside the
ISCO.  A second motivation for the present paper is therefore to
check the validity of the assumed line emissivity profile.

Our analysis is based on a non-relativistic hydrodynamic model of an
accretion disk. We present global numerical solutions of the
differential equations governing the fluid flow, assuming that the
accretion disk is steady, axisymmetric and in hydrostatic
equilibrium in the vertical direction, and using a pseudo-Newtonian
model for the gravitational potential. We do not include magnetic
fields explicitly, but assume an effective viscosity described by
the \al\ prescription (Shakura \& Sunyaev 1973). We also assume an
adiabatic index $\gamma=1.5$, which corresponds to approximate
equipartition between gas and magnetic pressure (Quataert \& Narayan
2000).

Our primary interest is in accretion disks in the rigorously defined
thermal state (see Table 2 in Remillard \& McClintock 2006) with $\HR
<0.1$, as these are the systems of most interest for our work on BH
spin (M06). Since the value of the viscosity parameter \al\ for such
disks is a matter of debate, we try different constant values: \al~=
0.01, 0.1, 0.2. We also consider a variable-\al\ prescription (eq. 22)
inspired by the MHD simulations of Hawley \& Krolik (2002, hereafter
HK02).  For non-spinning BHs, we use the pseudo-Newtonian potential of
Paczy\'nski \& Wiita (1980; hereafter PW80), and for spinning black
holes we use the pseudo-Kerr model of Mukhopadhyay (2002). The
numerical framework for our calculations is similar to that used by
Narayan, Kato \& Honma (1997), viz., we use a relaxation method to
solve the equations from the sonic radius \Rs\ to the outer edge of
the disk ($\sim 10^5$ \Rs), and we then integrate inward from \Rs\ to
the event horizon.

The paper is organized as follows.  We discuss in \S2 the theory and
computational method. We then discuss in  \S3 our numerical disk
solutions, focusing on the magnitude of the stress at the ISCO, the
amount of viscous dissipation near and inside the ISCO, and the
density and disk thickness contrast across the ISCO.
We then compute in  \S4 the
emitted spectra of our numerical disks for different values of $\HR$
and \al\ and investigate the error we make when we estimate the spin
of a BH via the continuum fitting method assuming the standard disk
model.  We conclude in  \S5 with a discussion.

\section
{THE MODEL}

\subsection{Gravity}

In order to focus our attention on the key physics of the problem,
and to avoid being distracted by technical details, we consider a
simple viscous hydrodynamic accretion disk in a Newtonian
gravitational potential.  Since the presence of an ISCO is essential
for our analysis, we simulate relativistic gravity in this Newtonian
model by means of a modified gravitational potential. For a
non-spinning BH, we make use of the PW80 potential:
\begin{equation}
  \Phi=-\frac{G M}{R-2 R_{\rm g}}\; ,
\end{equation}
where $M$ is the BH mass, $G$ the gravitational constant and $R_{\rm
g}=GM/c^2$. The Keplerian angular velocity \omk\ at a radius $R$
from the BH is
\begin{equation}
  \Omega_{\rm K}=\frac{ (G M)^{1/2}}{(R-2 R_{\rm g}) R^{1/2}}\; .
\end{equation}
In the case of a spinning BH we use the pseudo-Kerr model of
Mukhopadhyay (2002) in which the gravitational acceleration of a
test particle in a Keplerian orbit at a distance $R$ from the BH is
\begin{equation}
  F=- \nabla \Phi = \frac{c^4}{G M}\frac{(r^2- 2 a_* \sqrt{r} +
    a_*^2)^2}{r^3(\sqrt{r}(r-2)+a_*)^2}\; ,
\end{equation}
where $r=R/R_{\rm g}$, $a_*=a/M=J/(GM^2/c)$ is the dimensionless
spin of the BH, and $-1< a_*<1$. The Keplerian angular velocity at
radius $R$ is then
\begin{equation}
  \Omega_{\rm K}=\frac{c^3}{G M}\frac{(r^2 - 2 a_* \sqrt{r} +a_*^2)}{r^2
    (\sqrt{r} (r-2)+a_*)} \; .
\end{equation}

\subsection{Hydrodynamics}

We assume a steady axisymmetric disk in hydrostatic equilibrium in the
vertical direction. In the equations that follow, which have a long
history in accretion disk theory (e.g., Paczy\'nski \&
Bisnovatyi-Kogan 1981; Muchotrzeb \& Paczy\'nski 1982; Kato, Honma \&
Matsumoto 1988; Abramowicz et al. 1988; Popham \& Narayan 1991;
Narayan \& Popham 1993; Chen \& Taam 1993; Narayan et al. 1997; Chen,
Abramowicz \& Lasota 1997), we denote density, sound speed, radial
velocity, angular velocity, Keplerian angular velocity, and vertical
half-thickness by $\rho$, $c_{\rm s}$, $v_{\rm R}$, $\Omega$, \omk,
and $H$, respectively. All these parameters are taken to be functions
of the cylindrical radius $R$ only.  Because of the assumption of
steady state, the Lagrangian time derivative $D/Dt= \partial/\partial
t +\mathbf{v }\cdot \mathbf{\nabla}$ becomes $D/Dt=v_{\rm R}d/dR$.
After vertical and then radial integration the continuity equation
takes the form:
\begin{equation}
  4 \pi \rho v_{\rm R} R H=-\dot{M}={\rm constant}\; ,
\end{equation}
where $H= c_{\rm s}/\Omega_{\rm K}$. The momentum equation is
\begin{equation}
  \rho (\mathbf{v \cdot \nabla}) \mathbf{v}=-\mathbf{\nabla} P - \rho
  \mathbf{\nabla} \Phi
  +\rho \Omega^2 \mathbf{R} +\rho \mathbf{\nabla} \cdot
  \mathit{\sigma}\; ,
\end{equation}
where ${\it \sigma}$ is the stress tensor. We assume that the only
non-zero component of ${\it \sigma}$ is ${\it \sigma_{\rm {R
\Phi}}}= - \alpha P$ (\al\ prescription, Shakura \& Sunyaev 1973),
where $P$ is the total pressure and we write $P= \rho c_{\rm s^2}$.
The radial component of the momentum equation gives
\begin{equation}
  v_{\rm R} \frac{d v_{\rm R}}{d R}=-(\Omega_{\rm K}^2 -
  \Omega^2)R -\frac{1}{\rho} \frac{d}{d R}(\rho c_{\rm s}^2) \; ,
\end{equation}
and conservation of angular momentum gives
\begin{equation}
  \frac{\rho v_{\rm R}}{ R} \frac{d}{d R} (\Omega R^2)=
  \frac{1}{R^2 H} \frac{d(R^2 H \mathit{\sigma_{\rm {R
      \Phi}}})}{d R}\; .
\end{equation}
The latter equation can be integrated to obtain
\begin{equation}
  \Omega R^2 -j =-\frac{\alpha c_{\rm s}^2 R}{v_{\rm R}}\; ,
\end{equation}
where $\Omega R^2$ is the specific angular momentum of the gas at
radius $R$ and $j$ is an integration constant. We can interpret $j$ as
the specific angular momentum of the accreting gas at the radius where
the stress goes to zero.

Lastly, we write the energy conservation equation in terms of the
Lagrangian derivative of the specific entropy,
\begin{equation}
  \rho T \frac{D s}{Dt}= q^+- q^-= f q^+.
\end{equation}
Here $s$ is the specific entropy per unit mass, and $q^+$ and $q^-$
are the volume rate of heating and cooling of the gas, respectively.
Following Narayan et al. (1997) we take the cooling rate to be a
factor $(1-f)$ of the heating rate. Narayan et al. used $f=1$ because
they were modeling advection-dominated accretion flows. Since we are
interested primarily in thin disks, we use small values of $f$, i.e.,
substantial cooling, and we tune the value of $f$ to achieve the
desired disk thickness (eq. 19). The heating of the gas is due to
viscous dissipation, which gives $ q^+= \nu \sigma R
d\Omega/dR$. Using the relationship $\epsilon= P/(\gamma-1)$, where
$\epsilon$ is the thermal energy per unit volume and $\gamma$ is the
adiabatic index (we use $\gamma=1.5$), we can write
\begin{equation}
  \rho T \frac{D s}{Dt}= \frac{\rho v_{\rm R}}{\gamma-1}\frac{d c_{\rm
      s}^2}{dt}
  - c_{\rm s}^2 v_{\rm R} \frac{d \rho}{d R}\; .
\end{equation}
Thus, the energy equation takes the form
\begin{equation}
  \frac{\rho v_{\rm R}}{\gamma-1}\frac{d c_{\rm s}^2}{d R}- c_{\rm s}^2 v_{\rm R}
  \frac{d \rho}{d R}=-f\alpha \rho c_{\rm s}^2 R \frac{d \Omega}{dR}\; .
\end{equation}

\subsection{Boundary Conditions and Numerical Method}

We use a relaxation method to obtain numerical solutions of the above
differential equations.  In the computations, we define $x=R/R_{\rm
s}$ as the spatial variable and covered the region $x=1$ to $10^5$
using 1000 grid points. The grid has a non-uniform spacing, with more
grid points near the inner boundary $x=1$. In solving the equations,
we set $- 4 \pi \rho v_{\rm R} R H = \dot{M}=1$, and in order to
simplify the equations we substitute for $\rho$ using equation
(5). Thus we are left with three unknown functions of $R$: $v_{\rm
R}(R)$, $c_{\rm s}^2(R)$, and $\Omega(R)$. In addition, we have two
unknown constants, $j$ and $R_s$, which we treat as eigenvalues. To
solve for these quantities, we use equations (7), (9), and (12),
supplemented with five boundary conditions.

Narayan et al. (1997) showed that solutions of the disk model
described in \S2.2 tend to be nearly self-similar over a wide range of
radius.  Assuming self-similarity (following Narayan \& Yi 1994), we
can obtain the following analytic solution of the equations (the
subscript ``SS'' refers to self-similar):
\begin{equation}
c^2_{\rm s,SS}(R)= c_0^2 \frac{G M}{R}\; , \; c_0^2=\frac{2}{5+ 2
\epsilon'+ \alpha^2 / \epsilon'}\; , \;
\epsilon'=\frac{5/3-\gamma}{f(\gamma-1)}\; ,
\end{equation}
\begin{equation}
  v_{\rm R,SS} (R)= v_0 \sqrt{\frac{G M}{R}}\; ,\;
  v_0=-\alpha \sqrt{\frac{c_0^2}{\epsilon'}}\; ,
\end{equation}
\begin{equation}
\Omega_{\rm SS}(R)= \Omega_0 \Omega_K \; , \; \Omega_0=\sqrt{\frac{2
    \epsilon'}{5+ 2 \epsilon'+ \alpha^2/\epsilon'}}\; .
\end{equation}
We use this self-similar solution to set boundary conditions at the
outer boundary $R_{\rm
  out}$:
\begin{equation}
  v_{\rm R}(R_{\rm out})=v_0\sqrt{\frac{G M}{R_{\rm out}}}\; ,
\end{equation}
\begin{equation}
  c^2_{\rm s}(R_{\rm out})=c_0^2 \frac{G M}{R_{\rm out}}\; ,
\end{equation}
\begin{equation}
\Omega(R_{\rm out})=\Omega_0 \Omega_{\rm K}\;.
\end{equation}
From the above relations it can be shown that, at the outer
boundary, the vertical scale-height $H$ satisfies
\begin{equation}
  \frac{H}{R}=\sqrt{\frac{2}{5+2 \epsilon'+ \alpha^2/\epsilon'}} \; .
\end{equation}
Therefore, for a given value of $\gamma$, we can vary the disk
thickness $\HR$ by changing $f$. For $\gamma=1.5$, $f$= 0.000035 and
0.0035 give $H/R=0.01$ and 0.1, respectively.  Once set at the outer
edge, the value of $H/R$ remains constant over most of the disk,
becoming smaller only near and inside the ISCO.  Note that, for the
thin disk models that we consider in this paper which have $H/R \leq
0.1$, the advection parameter $f$ is very much less than unity.  This
means that radiative cooling (which is $\propto 1-f$) dominates by a
huge factor over energy advection ($\propto f$).  We briefly
discuss thicker advection-dominated solutions in \S5.

The inner boundary is at the sonic radius, $R=R_{\rm s}$, which is a
singular point of the differential equations. Following standard
methods, we obtain the following regularity conditions at $R_{\rm s}$:
\begin{equation}
  v_{\rm R}^2-\frac{2 \gamma}{\gamma+1}c_{\rm s}^2=0\; ,
\end{equation}
\begin{equation}
(\Omega_{\rm K}^2- \Omega^2)R- c_{\rm s}^2 \frac{2
    \gamma}{\gamma+1}\left(\frac{1}{R}-\frac{d
    \ln(\Omega_{\rm K})}{dR}\right)-c_{\rm s}^2
    \frac{\gamma-1}{\gamma+1}\frac{f
    \alpha R}{v_{\rm R}}\frac{d \Omega}{dR}=0.
\end{equation}
Equations (16)--(18), (20)--(21) provide the five boundary
conditions we need to find a unique solution. Once we have obtained
the solution between $R=R_{\rm s}$ and $R= R_{\rm out}$ via the
relaxation method, we use the solution at $R=R_{\rm s}$ as initial
conditions and integrate the equations from \Rs\ down close to the
BH event horizon.

We should emphasize that we do not set any boundary condition at
the ISCO.  Instead, we apply the boundary conditions at the sonic
radius, whose position is computed self-consistently for each
solution.  Further, even at the sonic radius, the viscous torque is
not set to zero --- the torque is computed self-consistently and is
allowed to continue smoothly inside the ISCO.  The numerical solutions
we obtain are thus superior to the standard disk model and can be used
to check the validity of the latter.  In particular, we can estimate
what error one makes in the standard disk model as a result of the
zero-torque boundary condition.

\section{RESULTS}

\subsection{Numerical Solutions}

Figure 1 shows model results for a non-spinning BH. We consider two disk
thicknesses: $\HR=0.01$ (solid lines), and $\HR=0.1$ (dotted lines). In
all four panels the vertical line shows the position of the ISCO ($R=6
R_{\rm g}$). We use $G=M=c=1$, so that the unit of velocity and time are
c and $G M/c^3$, respectively, and set $\dot{M}=1$. Most of our models
correspond to a constant value of \al. However, we also consider a model
in which \al\ varies as a function of $R$,
\begin{equation}
  \alpha= \frac{16.8}{(R/R_{\rm g})^3} +0.1 \; ,
\end{equation}
which closely reproduces the effective profile of \al\ found by HK02
(see their Fig. 4).  We refer to this as the ``variable-$\alpha$
model.''

Figure $1a$ shows the variation of the sound speed squared $c_{\rm
s}^2$ as a function of radius $R$.  For a given thickness, the
different \al\ models overlap at large radii and are only
distinguishable in the inner region of the disk. Here and in the
figures that follow, the magenta, blue, red and green lines refer to
the \al~= 0.01, 0.1, 0.2 and variable-\al~models, respectively.
Figure $1b$ shows the radial infall velocity {$v_{\rm R}$} of the
accreting gas.  We see that, between the ISCO and the event horizon,
{$v_{\rm R}$} increases rapidly regardless of the value of \al. The
variable-\al\ model almost completely overlaps with the \al = 0.1
model even at large radii. Figure $1c$ shows the angular velocity
$\Omega$ and Keplerian angular velocity \omk. The profiles of $\Omega$
for the different values of \al\ and $H/R$ are not distinct and are
represented by the single dotted line. The solid red line corresponds
to the Keplerian angular velocity. Note that the gas orbits in a
nearly Keplerian fashion until it reaches the ISCO.  Thereafter, the
hydrodynamic forces maintain an orbital motion that becomes
increasingly sub-Keplerian as the gas approaches the event
horizon. Figure $1d$ shows the gas density $\rho$ as a function of
radius. As in the case of the sound speed (Fig.  $1a$), the density
reaches a maximum outside the ISCO and then decreases rapidly near the
event horizon. In this plot, too, the variable-\al\ model coincides
with the \al~= 0.1 model.

Figure 2 is in the same format as Figure 1 and presents our results
for a spinning BH with $a_*=0.95$. The principal difference from the
previous figure is that the ISCO (vertical dashed line) is now located
at $R=1.937R_{\rm g}$. We consider the same values of $\alpha$ and
$H/R$ as in Figure 1, but there is no variable-$\alpha$ model in this
case because HK02 considered only a non-spinning BH.

\subsection{Matter Density, Disk Thickness and the Iron Line Method}

Before presenting our main results in the following subsections, we
briefly consider the implications of our models for the determination
of spin via the iron line method. The source geometry and illumination
law for producing the fluorescence iron line are probably the largest
uncertainties in the line fitting method (Reynolds \& Begelman
1997). If we assume the steepest law that is suggested by Reynolds \&
Begelman (1997), then the irradiating flux $F_{\rm X} \sim R^{-3}$.
Let us write the emissivity function in the form $ f_{\rm Fe} F_{\rm
X}$, where $f_{\rm Fe}$ is an efficiency factor. In this section we
investigate if the existing models of $f_{\rm Fe}$ in the literature
agree with our hydrostatic models.

The currently favored iron line models (BR06) assume that the iron
line emission is restricted between $R_{\rm ISCO}$ and an outer radius
$R_{\rm out}$ and that, within this region, the line profile is fitted
by a broken power law. BR06 find that the emissivity varies as $\sim
R^{-6}$ between the break radius $R_{\rm br}$ and $R_{\rm ISCO}$, and
as $\sim R^{-3}$ between $R_{\rm br}$ and $R_{\rm out}$.  For $F_{\rm
X} \sim R^{-3}$, this implies the following form for the efficiency
function:
\begin{equation}
\begin{array}{lll}
  f_{\rm Fe(BR06)}  & = 0 \; , & {\rm if }\; \: R < R_{\rm ISCO} \; ,\\
                    & = 1/R^3 \; , & {\rm if}\; \:R_{\rm ISCO}
                        \le R \le 3\:R_{\rm ISCO} \; , \\
                    & = {\rm constant}\; , & {\rm if }\;
                          \: 3\:R_{\rm ISCO} < R \; .
\end{array}
\end{equation}
Below we discuss the two main theories regarding the physical
parameters that might affect the emissivity profile.

Constant density models (Ross \& Fabian 2003; \.{Z}ycki et al. 1994;
Ross, Fabian \& Young 1999) predict that the line emissivity is
dependent on the ionization parameter, which is proportional to
$F_{\rm X}/\rho$, where $\rho$ is the gas density and $F_{\rm X}$ is
the illuminating flux. It is argued that the gas density drops to very
low values inside the ISCO. As a result, the region inside the ISCO
has a very high ionization parameter, which in turn produces
negligible iron line emission (Reynolds \& Begelman 1997; Young et
al. 1998; Fabian 2006).  In this case, one would expect $f_{\rm Fe}$
to be inversely related to ionization, i.e., $f_{\rm Fe}$ should be a
function of $\rho(R)/F_X(R) \propto \rho(R) R^3$ . More detailed
calculations that solve for the vertical structure of the disk under
hydrostatic equilibrium (e.g. Nayakshin et al. 2000, 2002) suggest
that the line emission depends on a ``gravity factor'' $\sim
(H/R^3)F_{\rm X}$. If that is the case, then for $F_{\rm X} \sim
R^{-3}$, one expects the efficiency function $f_{\rm Fe}$ to be
proportional to $H$.

In Figure 3, we compare the BR06 efficiency function $f_{\rm
Fe(BR06)}$ (eq. 23) to those suggested by our hydrostatic models,
in the context of the constant density and gravity theories mentioned
above.  Figure $3a$ shows $\rho(R) R^3$ as a function of radius. The
line types/colors for the various models are the same as those defined
in Figure 1.  Superimposed on our density profiles is a thick
short-dashed black line that represents $f_{\rm Fe(BR06)}$. For both
$\HR=0.01$ (solid lines) and $\HR=0.1$ (dotted lines), and all values
of \al, we note that $\rho(R) R^3$ is an increasing function of
radius, implying that $f_{\rm Fe}$ should also increase with
increasing radius. There is no apparent reason why $f_{\rm Fe}$ should
increase so steeply near the ISCO, or decrease at large radii, as
suggested by $f_{\rm Fe(BR06)}$.

Figure $3b$ shows a similar plot for a rapidly spinning BH with
$a_*=0.95$.  We notice the same trends as in Figure $3a$. In this case, we also
notice that (especially for $\HR=0.1$), instead of becoming negligible
at the ISCO, $\rho(R) R^3$ decreases gradually as one passes the ISCO
and moves closer to the event horizon. Therefore, one does not expect
$f_{\rm Fe}$ to drop abruptly to zero at the ISCO.

In Figures $3c$ and $3d$, we consider the disk thickness in the inner
region.  In our models, the disk has a more or less constant thickness
specified by $H/R$ outside $\sim 100 R_{\rm g}$, and we vary this
``outer thickness'' by changing the value of $f$ (eq. 19).  However,
in the inner region, the disk gets thinner.  Figure $3c$ shows $H$ as
a function of $R$ for a non-spinning BH.  The top panel shows a disk
with outer thickness of 0.01 and the bottom one shows a disk with
outer thickness of 0.1 for the choices of $\alpha$ specified in \S3.1.
In the thinner case, there is an abrupt drop in $H$, which would
likely quench the iron emission from inside the ISCO.  For the thicker
case, however, the value of $H$ decreases gradually and remains
significant far inside the ISCO at $3R_{\rm g}$.  Thus, these models
indicate that the region within the ISCO may contribute a significant
fraction of the total iron line emission and, also that it is
difficult to justify the steeply falling form of $f_{\rm
Fe(BR06)}$. As shown in Figure $3d$, the results for a BH with
$a_*=0.95$ ($R_{\rm ISCO} = 1.973R_{\rm g}$) are very similar. Again,
for $H/R = 0.1$ the disk thickness $H$ decreases gradually near and
within the ISCO.

We hasten to add that this is a very simple model of an accretion
disk, perhaps too simple to address ``surface phenomena'' such as
fluorescent iron line emission.  Modulo this important caveat it seems
that, for reasonable values of the model parameters, the iron line
emission does not necessarily end at the ISCO, nor does it vary with
radius outside the ISCO with anything like the functional form assumed
in current fits of iron line data (e.g., BR06).

\subsection{Viscous Stress Near the ISCO}

Figure $4$ shows the vertically integrated stress $2 H \alpha P$ for a
non-spinning BH.  As shown in Figure $4a$, all the models
corresponding to a very thin disk are in close agreement with the
standard model, i.e., the stress nearly vanishes at the ISCO even
though we do not require this of the model. For the thicker disk
shown in Figure $4b$, the stress near and inside the ISCO increases,
the effect becoming more important for larger values of \al.
Interestingly, for $\alpha=0.01$, the  magnitude of the peak
stress is actually smaller than that predicted by the standard disk
model.

As shown in Figure $5$, our models for a spinning black hole display
essentially this same dependence of stress on $H/R$ and \al.  In both
Figures 4 and 5, the presence of a non-zero viscous stress inside the
ISCO implies a contribution to the observed spectrum that is not
accounted for in the standard disk model. In \S4 we investigate the
magnitude of this effect.

We now consider the effect of \al\ and disk thickness on the
eigenvalue $j$ (\S2.2), which is the specific angular momentum
delivered to the black hole by the infalling matter.  In the standard
disk model, $j$ is the Keplerian specific angular momentum at the ISCO
because (i) matter is assumed to orbit at the Keplerian velocity and
(ii) the stress is assumed to vanish inside the ISCO.  Neither
assumption is made in our hydrodynamic models, and it is therefore of
interest to consider how much the calculated values of $j$ differ from
the standard value.  Table~1 summarizes the values of $j$ for our
different models. We note that $j$ decreases with increasing $\HR$ and
\al.  That is, as the disk gets thicker or as \al\ increases, more
angular momentum is removed before matter falls into the BH.  However,
the effects are quite small, and the deviations are less than 1\% in
all cases.

\subsection{Dissipation Inside The ISCO}

In the previous section we showed that the stress at the ISCO is small,
but non-zero, and that it increases with disk thickness and \al.  We now
consider the energy dissipation profiles of our model disks for
different values of \al\ and $\HR$. Figure $6$ shows the quantity,
\begin{equation}
  R \frac{dL}{dR}=\frac{dL}{d \ln (R)}=\frac{-\dot{M}\alpha c_{\rm
  s}^2 R^2}{v_{\rm R}} \frac{d\Omega}{dR}= 4 \pi R^2 D(R) \,,
\end{equation}
as a function of $R$. Here $L$ is the luminosity and $D(R)$ the energy
dissipated per unit time per unit surface area of the disk.  Figures
$6a$ and $6b$ show $R dL/dR$ vs $R$ for $a_*=0$, while Figures $6c$
and $6d$ show the results for $a_*=0.95$.  The solid black lines show
the standard disk model with zero torque at the ISCO.  For the thin
disk with $\HR =0.01$ and for all values of \al, our models are
indistinguishable from the standard model, which thus provides an
excellent description of the flow in this case.  However, for the
thicker disk with $\HR=0.1$, our numerical models deviate somewhat
from the standard disk model.  We note in particular that larger
values of \al\ are associated with more dissipation near the ISCO and
larger deviations from the standard disk model.

In Table 1 we summarize the total luminosities of the different models
for a given mass accretion rate \mdot.  We note that none of the
luminosities of our models deviates by more than 4\% from that of the
standard model.

\begin{center}
  \begin{tabular}{ccccccc}
    \multicolumn{7}{c}{Table 1} \\ & \\
    \multicolumn{7}{c}{Luminosities and Angular Momentum Eigenvalues
                       of the Numerical Disk Models} \\ \\
    \hline
    \hline
    \multicolumn{1}{c}{$a_*$}&\multicolumn{1}{c}{H/R}&\multicolumn{1}{c}
    {$\alpha$}&\multicolumn{1}{c}{$L_{\rm TOTAL}$}
    &\multicolumn{1}{c}{$L_{\rm TOTAL(STD)}$}&\multicolumn{1}{c}{$
    j$}&\multicolumn{1}{c}{$j_{\rm STD}$}\\
    \multicolumn{1}{c}{}&\multicolumn{1}{c}{}&\multicolumn{1}{c}{}&\multicolumn{1}{c}{${\rm
    (\dot{M}c^2)}$}
    &\multicolumn{1}{c}{${\rm (\dot{M} c^2)}$}&\multicolumn{1}{c}{}
    &\multicolumn{1}{c}{} \\
    \hline
    0&0.01 & 0.01 & 0.0624    & 0.0625  & 3.6744 & 3.6742   \\
    &    & 0.1   & 0.0625     &         & 3.6735 &          \\
    &    & 0.2   & 0.0626     &         & 3.6727 &          \\
    &    & variable & 0.0626  &         & 3.6730 &          \\
    \hline
    &0.1  &0.01 &  0.0610     &         & 3.6839 &          \\
    &    &0.1  &  0.0633      &         & 3.6609 &          \\
    &    &0.2  & 0.0650       &         & 3.6456 &          \\
    &    &variable&0.0646     &         & 3.6518 &          \\
    \hline
    0.95 & 0.1&0.01 & 0.2091  &  0.2144 & 2.3372   & 2.3311 \\
    &    &0.1       &    0.2171 &       &  2.3237  &    \\
    &    &0.2       &  0.2227   &       &  2.3146  &     \\

    \hline 
    \multicolumn{7}{l}{The subscript STD refers to the standard thin
    disk model.} \\

  \end{tabular}
\end{center}

\section{DISK SPECTRA AND THE EFFECT ON BH SPIN ESTIMATION}

In the standard disk model, the viscous dissipation is assumed to vanish
at the ISCO.  As a result, the emitted flux also vanishes at the ISCO,
and no radiation is emitted from the region of the flow between the ISCO
and the event horizon.  For a given BH mass, the radius of the ISCO is a
well-known and monotonically decreasing function of $a_*$, e.g., for
$a_*$= 0, 1, the ISCO is located at $ 6 R_{\rm g}$, $1 R_{\rm g}$,
respectively.  As discussed in Zhang et al. (1997), S06 and M06, the
radius $R_{\rm in}$ of the inner edge of the disk can be estimated from
observations.  For a BH of known mass, this radius can be expressed in
units of $R_{\rm g}$, and if the disk inner edge is located at the ISCO,
then $R_{\rm in}/R_g$ determines the spin parameter $a_*$.

From the calculations presented in this paper, we see that for a
very thin disk ($\HR=0.01$) the viscous dissipation does indeed
become negligible inside the ISCO and the dissipation profile $R
dL/dR$ is identical to that predicted by the standard disk model.
Thus for such systems we expect our estimates of BH spin to be quite
accurate. However, we do notice a difference for thicker disks with
say $H/R \sim0.1$. Using the standard disk model to fit the observed
spectra of these systems will lead to an error in our estimate of
the radius of the ISCO.  We now try to quantify this error.

For each of our disk solutions, we have calculated the emitted
spectrum assuming that the disk emits like a blackbody at each
radius. The temperature profile $T(R)$ of the disk surface can be
calculated from $dL/dR$ using:
\begin{equation}
  (1-f)\frac{dL}{dR}= 4\pi \sigma R T^4(R) \; ,
\end{equation}
where $\sigma$ is the Stefan-Boltzmann constant. This can be used to
calculate the observed spectrum of the disk by integrating over the
entire disk:
\begin{equation}
  F_{\rm \nu}= \frac{2 \pi \cos i}{D^2}\int_{R_{\rm inner}}^{R_{\rm
      out}} \frac{2 h
    \nu^3 R dR}{c^2 e^{\left( h\nu/ k T(R)) -1\right )}} \; ,
\end{equation}
where $h$ is the Planck constant, $c$ the speed of light, $k$ the
Boltzman constant, $D$ the distance, $i$ the angle of inclination,
$\nu$ the frequency, and $R_{\rm {inner}}$ the radius of the inner
boundary of the disk, near the event horizon.

Figure $7$ shows our calculated spectra for a BH with mass $M=10
M_{\odot}$ and distance $D=10$ kpc. In each panel the solid curve
shows the spectrum from a standard disk model with the appropriate
pseudo-Newtonian potential.  As before, we have considered three
constant values of $\alpha$: 0.01, 0.1 and 0.2, for both the spinning
and non-spinning cases, and an additional variable-\al\ model for the
non-spinning case.

Figures $7a$ and $7b$ show the calculated spectra for the case of a
non-spinning BH.  For $H/R=0.01$, we see that the calculated spectra
for all four models of \al\ overlap with the spectrum calculated via
the standard disk model. Therefore, we can conclude that, for such
very thin disks, the standard disk model is a very good approximation
and that the choice of \al\ cannot be a major source of error in
estimating BH spin. The different curves are more distinct in the case
of a thicker disk with $\HR=0.1$ (magenta, blue, red and green lines
show the \al~= 0.01, 0.1, 0.2, and variable-\al\ models). The
differences are especially noticeable at high photon energies, where
larger values of \al\ give higher fluxes. Figures $7c$ and $7d$ show
$H/R=0.1$ disk spectra for $a_*=0.8$ and 0.95.

To estimate how the spectral distortions might affect BH spin
determination, we produced spectral data files for our models using
an $RXTE$ response file and analyzed the data with XSPEC {\it
version 12.2.0}. These ``fake'' data files were fitted with the
XSPEC model $Diskpn$ (Gierlinski et al. 1999), which uses the
standard disk model with the PW80 potential and a zero torque
boundary condition at the ISCO. $Diskpn$ has three fit parameters:
\Tmax, $R_{\rm in}/R_{\rm g}$ and normalization $K=M^2\cos
i/D^2\beta$, where $M$ is the mass, $D$ the distance, $i$ the angle
of inclination, and $\beta$ the color correction factor. We are
interested in the case when the inner edge of the disk coincides
with the ISCO. Thus, since $Diskpn$ considers a non-spinning BH, we
set $R_{\rm in}= 6 R_{\rm g}$. The constant $K$ can then be
rewritten as
\begin{equation}
  K=\left( \frac{R_{\rm in}}{8.86 \times 10^6 \, {\rm cm}} \right )^2
  \left ( \frac{D}{10 \, {\rm kpc}} \right )^{-2} \frac{\cos i}{\beta} \; .
\end{equation}
In the above expression, $8.86 \times 10^6$ cm corresponds to $6 R_{\rm
g}= R_{\rm ISCO}$ for a non-spinning black hole with $M= 10
M_\odot$. From the value of $K$ obtained from spectral fitting, one can
calculate $R_{\rm in}$ for each model using equation (27). Using this
value of $R_{\rm in}$, one can then calculate the BH spin for which the
ISCO would be located at that radius. This is the spin that one infers
from the fake spectral data, under the assumption that the standard disk
model is correct.  Since the model was calculated with full viscous
hydrodynamics as described in earlier sections, the spin value derived
assuming the standard disk model will be different from the true BH spin
($a_*=0$ in this case).  The difference between the two values
represents the error in the spin estimate, $\Delta a_*$, caused by our
use of the simplified standard disk model.

The results of this analysis are summarized in Table 2.  The results
correspond to $M= 10 M_\odot$, $D$= 10 kpc, $\cos i=1$ and
$\beta=1$.

\begin{center}
  \begin{tabular}{cccc}
    \multicolumn{4}{c}{Table 2} \\ & \\
    \multicolumn{4}{c}{Errors in Spin Estimation of a Non-spinning BH} \\ \\
    \hline
    \hline
    \multicolumn{1}{c}{$H/R$}&\multicolumn{1}{c}{$\alpha$}&\multicolumn{1}{c}{$R_{\rm
    in}/(8.86 \times 10^6 \, {\rm cm})$}&\multicolumn{1}{c}{$\Delta a_*$} \\
    \hline
    0.01 & 0.01  &1.011  &-0.011  \\
    & 0.1 & 1.008  & -0.008  \\
    & 0.2 & 1.006  & -0.006  \\
    & variable& 1.007 & -0.004 \\
    \hline
    0.1  &0.01 & 1.021   &  -0.019 \\
    &0.1  & 0.960 &    0.037 \\
    &0.2  & 0.920  &   0.074  \\
    &variable&0.879 &  0.060 \\

    \hline
  \end{tabular}
\end{center}

Figure $8a$ shows the results in more detail for $\HR=0.01$, 0.02, 0.04,
0.06, 0.08 and 0.1. Figure $8b$ shows the results for the variable-\al\
model as a function of disk thickness. We see that the error is larger
for thicker disks and also for larger values of \al.  However, even for
the thickest case we considered, $\HR=0.1$, and the largest value of
$\alpha=0.2$, the BH spin is overestimated by less than $0.1$.  Thus, in
the case of a non-spinning BH the error is quite modest when one
considers, for example, that both the radius of the ISCO and the binding
energy at the ISCO differ only slightly (by 6\%) for a BH with $a_* =
0.1$.
%We also notice that for $\HR=0.01$ the spin has a
%small non-zero negative value.
%This effect
%is not visible in the figures, where all the different \al\ models
%seem to coincide with the standard disk model.

Though a similar standard pseudo-Kerr $XSPEC$ model is not available
for fitting our spinning BH model spectra, it is still possible to
estimate $\Delta a_*$ by calculating the model luminosities.
Gierlinski et al. (1999) showed that, for a non-spinning BH with the
PW80 potential, one can write:
\begin{equation}
  L=\frac{1}{16} \dot{M}c^2 =35.7 \frac{\pi
    \sigma}{\beta^4}R_{\rm in}^2 T_{\rm max}^4  \; ,
\end{equation}
where $L$ is the luminosity, $\sigma$ the Stefan-Boltzmann constant,
$\beta$ the color correction factor, $R_{\rm in}$ the radius of the
inner edge of the disk, and $T_{\rm max}$ the peak temperature of
the disk. Therefore, instead of calculating the multicolor blackbody
spectrum of our models and fitting them with XSPEC, we could simply
compare each hydrodynamic model with the corresponding standard disk
model with the same $T_{\rm max}$.  This gives the following
estimate for the effective disk inner radius $R_{\rm in}$ of any
given hydrodynamic model,
\begin{equation}
  R_{\rm in}^2= \frac{ L_{\rm model}}{L_{\rm standard \; disk}}
  R_{\rm ISCO}^2 \; ,
\end{equation}
where $L_{\rm model}$ is the luminosity of the model, $L_{\rm
standard \; disk}$ is the luminosity of the standard disk with the
same value of $T_{\rm max}$, and $R_{\rm ISCO}$ is the radius of the
ISCO. The value of $R_{\rm in}$ obtained using equation (29) may
then be used to calculate $\Delta a_*$, as before.

Figure 9a shows $\Delta a_*$ values calculated using both the full
spectral fitting method via equation (27) and the simpler
luminosity-temperature method described by equation (29). We see that
the results are very close, indicating that the second method is a
good proxy for the more detailed spectral method.

For a spinning BH, equation (28) can be generalized to
\begin{equation}
  L=\epsilon \dot{M}c^2 \sim c_0 \frac{\pi \sigma}{\beta^4}R_{\rm in}^2
T_{\rm max}^4 \; ,
\end{equation}
where $\epsilon$ is the spin-dependent efficiency of the BH, and
$c_0$ is a constant. Therefore, equation (29) can again be used to
estimate the effective $R_{\rm in}$ and this can be used to obtain
an estimate of the BH spin.

Figure $9b$ shows $\Delta a_*$ for spinning BHs using this method.
We show results for $a_*$ = 0.7, 0.8, 0.9 and 0.95, and $\HR=0.1$.
For a given disk thickness and \al, we see that the error in the
spin estimate becomes smaller as the spin of the BH increases. For
$a_*=0.95$, the maximum error is only $\sim 0.01$.

\section{DISCUSSION}

In this paper we studied the properties of a simple hydrodynamic model
of an accretion disk using the $\alpha$ prescription for viscosity. We
considered models with finite thicknesses $H/R$ and different values
of $\alpha$.  Our aim was to investigate how much the hydrodynamic
models of thin disks deviate from the idealized ``standard disk
model'' which assumes a vanishing torque at the innermost stable
circular orbit (ISCO).

We find that the deviations of the viscous hydrodynamic models from
the standard disk model increase with increasing $H/R$ and increasing
$\alpha$.  However, even for $H/R=0.1$ and $\alpha~= 0.2$, the largest
values we tried for our thin disk calculations, the deviations remain
modest. This is illustrated in Figures 4 and 5, which show how the
stress profile deviates from that of the idealized standard disk
model, and also in Figure 6, which compares the profiles of the
viscous energy dissipation rate $R dL/dR$, Figure 7, which shows the
multicolor blackbody spectra of the models, and Table 1, which gives
some quantitative results.  In all cases, we see that the detailed
hydrodynamic models match the standard disk model quite closely.

We were motivated to do this study because we and others have used the
standard disk model to fit the continuum spectra of BH X-ray binaries
in the thermal state in order to estimate the spins of the BHs.  How
much error do we expect in the estimated spin values as a result of
the fact that a real disk deviates from the standard disk model?  At
least for the simple hydrodynamic models we have considered in this
paper, the answer is that the errors are quite modest.

Quantitative results are given in Table 2 and Figures 8 and 9.  The
error $\Delta a_*$ in the derived estimate of BH spin is at most
$\sim0.1$ in the case of a non-spinning BH and is much less for
rapidly spinning BHs.  These errors are comparable to or smaller
than the errors that arise from uncertainties in our estimates of
mass, distance and disk inclination (S06, M06).

While these results are very encouraging for our program to estimate
BH spin through fitting the continuum spectra of BH accretion disks in
the thermal state, we must note some caveats.  First and foremost, we
have considered a highly simplified toy hydrodynamic model with an
$\alpha$ prescription for viscosity.  Real disks doubtless have
magnetic fields, and the stresses associated with these fields
probably do not behave like microscopic viscosity.  Indeed, it is
precisely this argument that has been used  by Krolik (1999),
Gammie (1999) and HK02 to question the zero-torque boundary condition
at the ISCO.  On the other hand, Paczy\'nski (2000) makes an equally
persuasive argument (based on the angular momentum conservation
equation) that, so long as the shear stress is smaller than the
pressure, a thin disk will always satisfy the zero-torque condition.

In an attempt to include some of the effects of magnetic fields, we
have considered a model in which we allowed $\alpha$ to vary with
radius (see eq. 22) in such a manner as to closely mimic the effective
$\alpha$ obtained by HK02 from their MHD simulations.  Even though in
this model $\alpha$ increases rapidly with decreasing radius,
especially inside the ISCO, we found that none of our results changed.
Based on this finding we cautiously suggest that the inclusion of
magnetic fields may not significantly alter our conclusions.

One question that needs to be addressed is why our results differ so
much from those obtained by HK02.  From MHD simulations of magnetized
gas accreting in a PW80 potential, those authors concluded that the
vertically integrated magnetic stress increases monotonically with
decreasing radius all the way through and inside the ISCO.  This is
dramatically different from the behavior we find, as a comparison of
HK02's Fig. 10 with our Fig. 4 shows.  A likely explanation is that we
have limited our study to {\it thin} disks ($H/R=0.01, ~0.1$) in which
we simulated strong cooling by choosing a small value for the
advection parameter $f$ (see the discussion below eq. 19).  HK02, by
contrast, had no cooling in their MHD simulation, so their gas
retained whatever energy was generated through shocks, making their
disk thicker.

In order to verify that this difference is important, we calculated
models with larger values of $f$ using our viscous hydrodynamic code.
It is hard to know what effective value of $f$ is most appropriate to
match the HK02 simulation.  Nominally, theirs was a fully
advection-dominated accretion flow, since they had no cooling at all;
this means that their simulation corresponded to $f=1$.  However, we
do not know how well their code conserved energy.  Therefore, we
calculated three models with $f=1$, 0.5 and 0.25, all with the
variable $\alpha$ prescription (eq. 22) which most closely matches
their stress profile.  Figure 10 shows the resulting stress profiles.
We see that these advection-dominated models do exhibit a
monotonically increasing stress inward, exactly as found by HK02
(their Fig. 10).  The stress profiles are very different from those we
find for cooling-dominated thin disks (our Figs. 4 and 5).  Thus, we
tentatively suggest that a large part of the difference between the
results we find in this paper and those obtained by HK02 is related to
the differing treatments of the energy equation of the gas, viz.,
cooling-dominated thin disk regime versus advection-dominated thick
disk regime.  In other words, we confirm the original insight of
Abramowicz \& Kato (1989), Paczy\'nski (2000) and Afshordi \&
Paczy\'nski (2003) on the strong relation between disk thickness and
the stress at the ISCO.  However, only a detailed MHD study of an
accretion disk with significant cooling can tell for sure if this
interpretation is correct, and to our knowledge nobody has carried out
such a study.

Another limitation in our work is that we used a Newtonian model and
we simplified the thermodynamics of the gas in the disk via the
advection parameter $f$ (see eq. 10).  However, doing the calculations
in general relativity with full radiation thermodynamics will, we
believe, introduce modifications only of order unity.  The changes
will be larger for a spinning BH, which we modeled with the
Mukhopadhyay (2002) model, compared to a non-spinning hole (PW80
potential), but we think the error will still be only of order unity.
Therefore, calculating these effects in more detail will not greatly
alter our qualitative conclusion that the standard disk model is
adequate so long as the disk geometrically thin.  Nevertheless,
it would be useful to extend this work using a more complete set of
disk equations, such as those employed in the study of slim disks
(Abramowicz et al. 1988), and with the inclusion of general
relativity (e.g., Abramowicz, Lanza \& Percival 1997).

Note that we employed the two pseudo-Newtonian potentials mentioned
above in the work reported here merely to obtain a ballpark estimate
of the error associated with the zero-torque approximation.  When we
actually fit data to estimate the spin parameters of BHs (e.g., S06,
M06), we use a detailed model (Li et al. 2005) which assumes the Kerr
metric and includes all special relativistic and general relativistic
effects.

In our work on BH spin (S06, M06), we limited ourselves to disks with
luminosities less than 30\% of Eddington, which corresponds to
vertical thicknesses $H/R<0.1$.  The present study shows that this was
a reasonable choice.  For $H/R \leq 0.1$, the effects of gas physics
and finite vertical thickness in our hydrodynamic models are not
serious.  Equally clearly, for thicker disks with $H/R$ much greater
than 0.1, the effects will be large; e.g., see Figure 10.  Therefore,
one should be cautious about applying the standard disk model to disks
more luminous than 30\% of Eddington.  For this reason, we
believe the results obtained by Middleton et al. (2006) for the spin
of the microquasar GRS 1915+105 should be taken with caution.

Strong observational evidence that fitting the X-ray continuum is a
promising way to estimate black hole spin comes from a long history of
fitting the broadband spectra of black hole transients using the simple
non-relativistic multicolor disk model (Mitsuda et al. 1984; Makishima
et al. 1986), which returns the temperature $T_{\rm in}$ at the
inner-disk radius $R_{\rm in}$.  In their classic review, Tanaka \&
Lewin (1995) give examples of the steady decay (by factors of 10--100)
of the thermal flux of transient sources during which $R_{\rm in}$
remains constant.  They remark that the constancy of $R_{\rm in}$
suggests that it is related to the radius of the ISCO.  More recently,
this evidence for a constant inner radius in the thermal state has been
presented for a number of sources via plots showing that the bolometric
luminosity of the thermal component is approximately proportional to
$T_{\rm in}^4$ (Kubota \& Makishima 2001; Kubota \& Makishima 2004; Abe
et al. 2005; McClintock et al. 2007).  In short, these non-relativistic
analyses, which ignore spectral hardening (Davis et al. 2006), provide
evidence for the presence of a stable radius, although they obviously
cannot provide a secure value for the radius of the ISCO or even
establish that the stable radius is the ISCO.

We now consider the iron-line method of estimating spin. In this
method, it is assumed that the line emission ceases abruptly at the
ISCO, so an important question is whether or not the gas inside the
ISCO will fluoresce (Reynolds \& Begelman 1997). The possibility of
line emission from inside the ISCO is usually discounted on the
grounds that the density will fall suddenly inside the ISCO, thus
causing a sudden increase in the ionization parameter (Fabian 2007,
and references therein). Alternatively, and to the same effect, it is
argued that emissivity is related to the ``gravity parameter''
(Nayakshin 2000, 2002), and should depend on $H$. We see in Figure
$3a$ that the density dependent function $\rho(R) R^3$ does become
negligible inside the ISCO for the non-spinning BH. However, that is
not the case for a fast spinning BH with $a_*=0.95$ (Figure $3b$) even
for a small disk thickness of $\HR=0.1$. Also, the radial dependence
of $H$ shown in Figures $3c$ and $3d$ implies that there should be
emission from the inner region unless the disk is very thin
($\HR=0.01$).

An additional complication for iron line modeling is that the
emissivity is assumed to vary as a broken power-law, with the maximum
emission occurring exactly at the ISCO (e.g., BR06).  Looking at
Figure 3, such an ad hoc model would be hard to justify if the
emissivity has anything to do with gas density or disk thickness.  In
contrast, the continuum-fitting model has the merit that it makes use
of a physically motivated profile of disk emission $R dL/dR$ which can
be calculated from first principles in the standard disk model and
which continues to be valid even in the more general hydrodynamic
models described in this paper (Figs. 6, 7).
%It would thus appear
%that, at the present time, the continuum fitting method is the more
%robust approach to estimating BH spin.  Apart from being grounded better
%in physics, the method seems reliable up to $H/R \sim 0.1$, whereas the
%iron line method probably becomes unreliable for much thinner disks.

This paper has focused on only one aspect of BH spin estimation, viz.,
the validity of assumptions made in various methods of spin
determination regarding the hydrodynamical properties of the accretion
disk.  Of course, a successful determination of spin needs more than a
valid disk model.  It also requires high quality data and accurate
determination of secondary system parameters.  A discussion of these
issues is beyond the scope of this paper, and the reader is referred
to appropriate papers in the literature (e.g., M06; BR06).

\acknowledgements The authors thank Niayesh Afshordi and Jonathan
McKinney for discussions and useful suggestions.  We dedicate this
paper to Bohdan Paczy\'nski for his amazing insights in accretion
theory and in numerous other areas of astrophysics.

\newpage
\begin{figure}
  \figurenum{1} \plotone{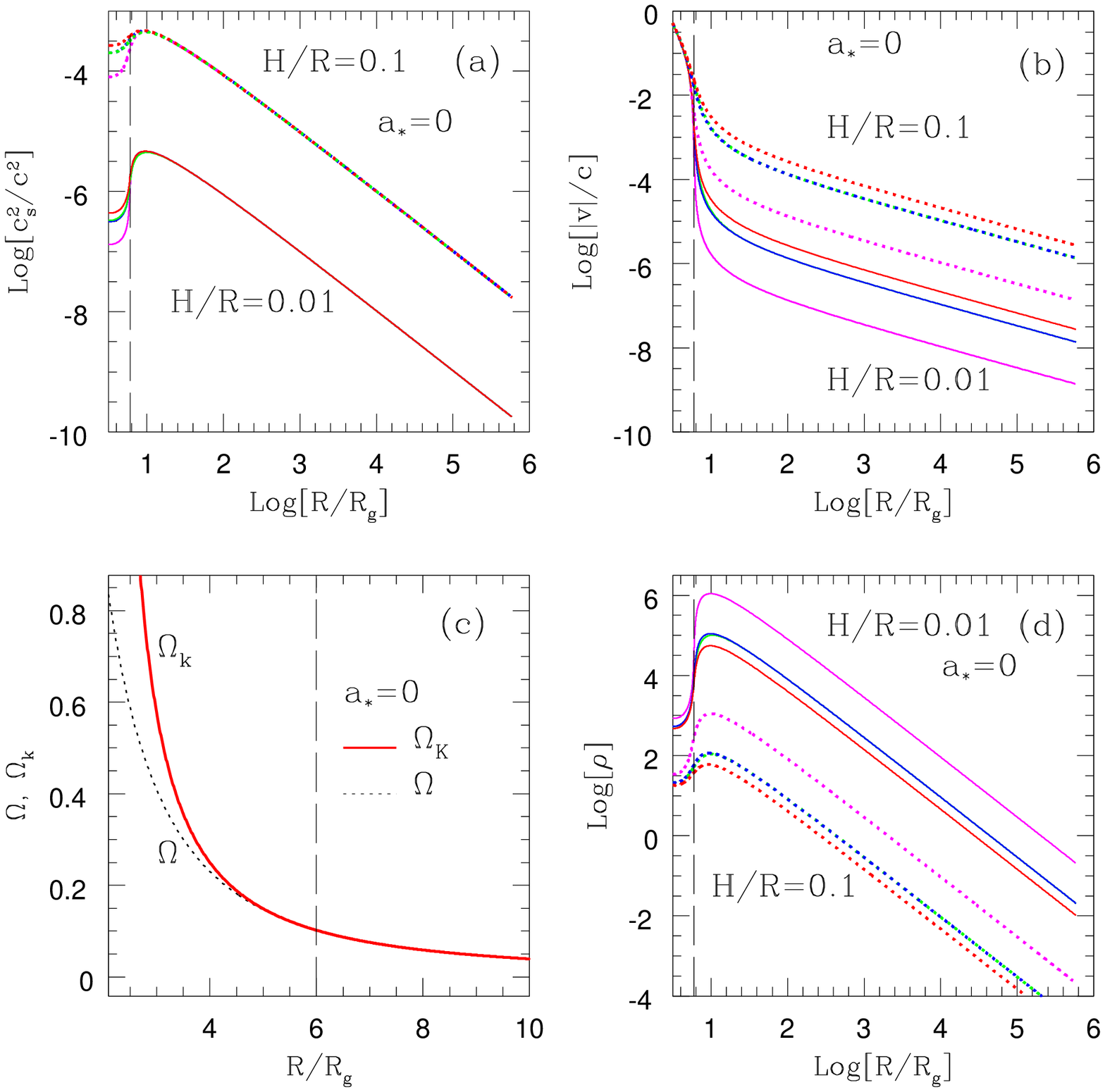} \caption{ Disk parameters for a
  non-spinning BH: ($a$) sound speed, ($b$) radial velocity, ($c$)
  angular velocity, ($d$) density. In panels $a$, $b$, and $d$, the
  solid and dotted lines correspond to $\HR=0.01$ and $\HR=0.1$,
  respectively; where distinct, the magenta, blue, red and green lines
  represent $\alpha = 0.01$ , 0.1 and 0.2, and variable \al\ (eq. 22),
  respectively. In all three panels, the variable-\al~model is nearly
  coincident with the \al~= 0.1 model.  In panel $c$, the Keplerian
  velocity is plotted as a solid red line.  Because the angular-velocity
  profile of the four models are nearly identical, we represent them by
  a single dotted line.  The radius of the ISCO, $R=6R_{\rm g}$, is
  indicated in all four panels by the vertical dashed line.  All
  numerical values correspond to $G=c=M=1$, $\dot{M}=1$. In panel $c$,
  the unit of angular velocity is $(G M/c^3)^{-1}$. In panel $d$, the
  unit of $\rho$ is $c^6/(G^3 M^2)$.}
\end{figure}

\newpage
\begin{figure}
  \figurenum{2} \plotone{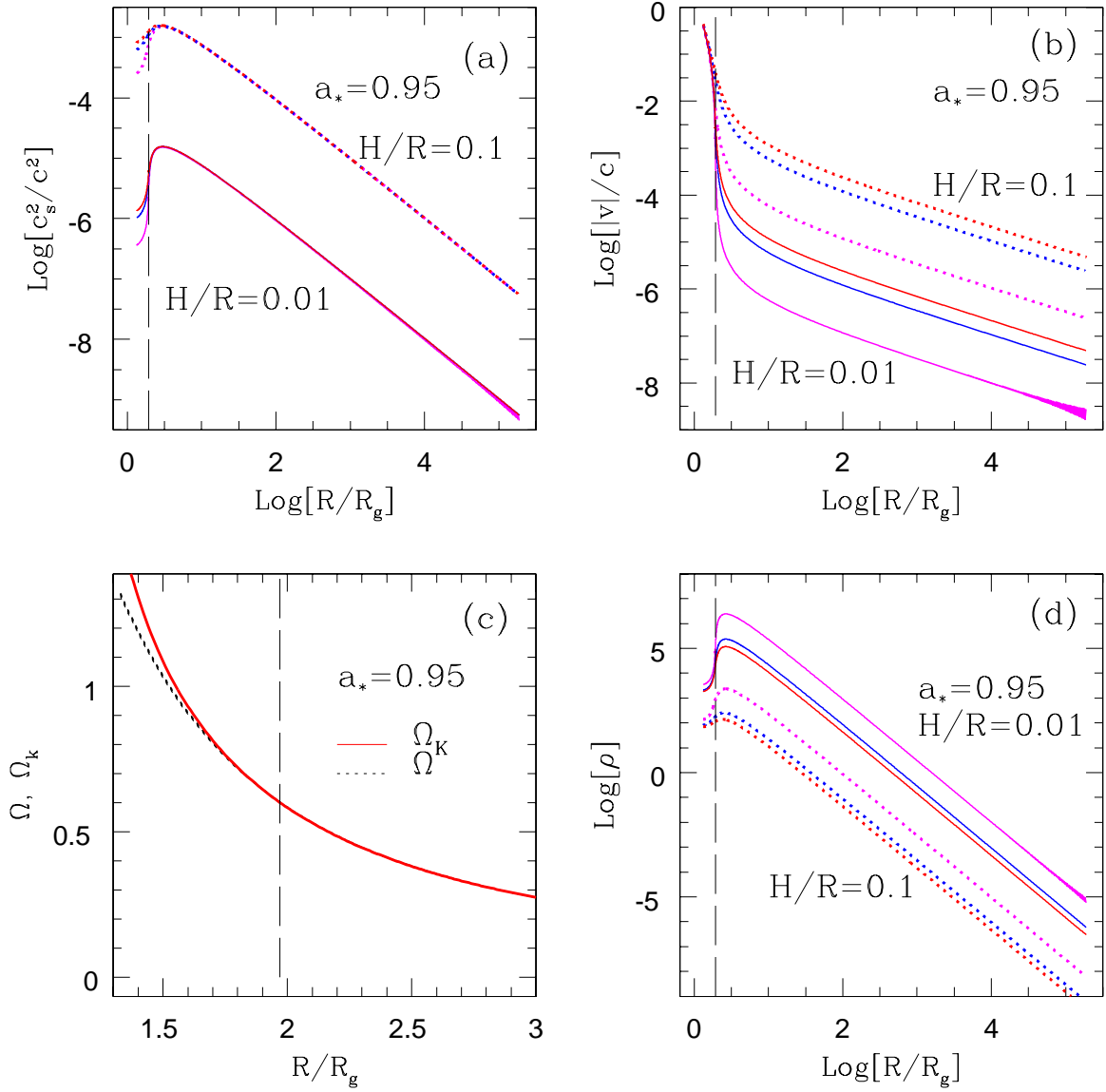} \caption{Similar to Figure 1, but for
    a spinning BH with $a_*=0.95$. The vertical dashed line shows the
    ISCO at R= $1.937 R_{\rm g}$.  There is no variable-\al~model in
    this case (see text), and hence the green line is not present in
    these plots.}
\end{figure}

\newpage
\begin{figure}
  \figurenum{3} \plotone{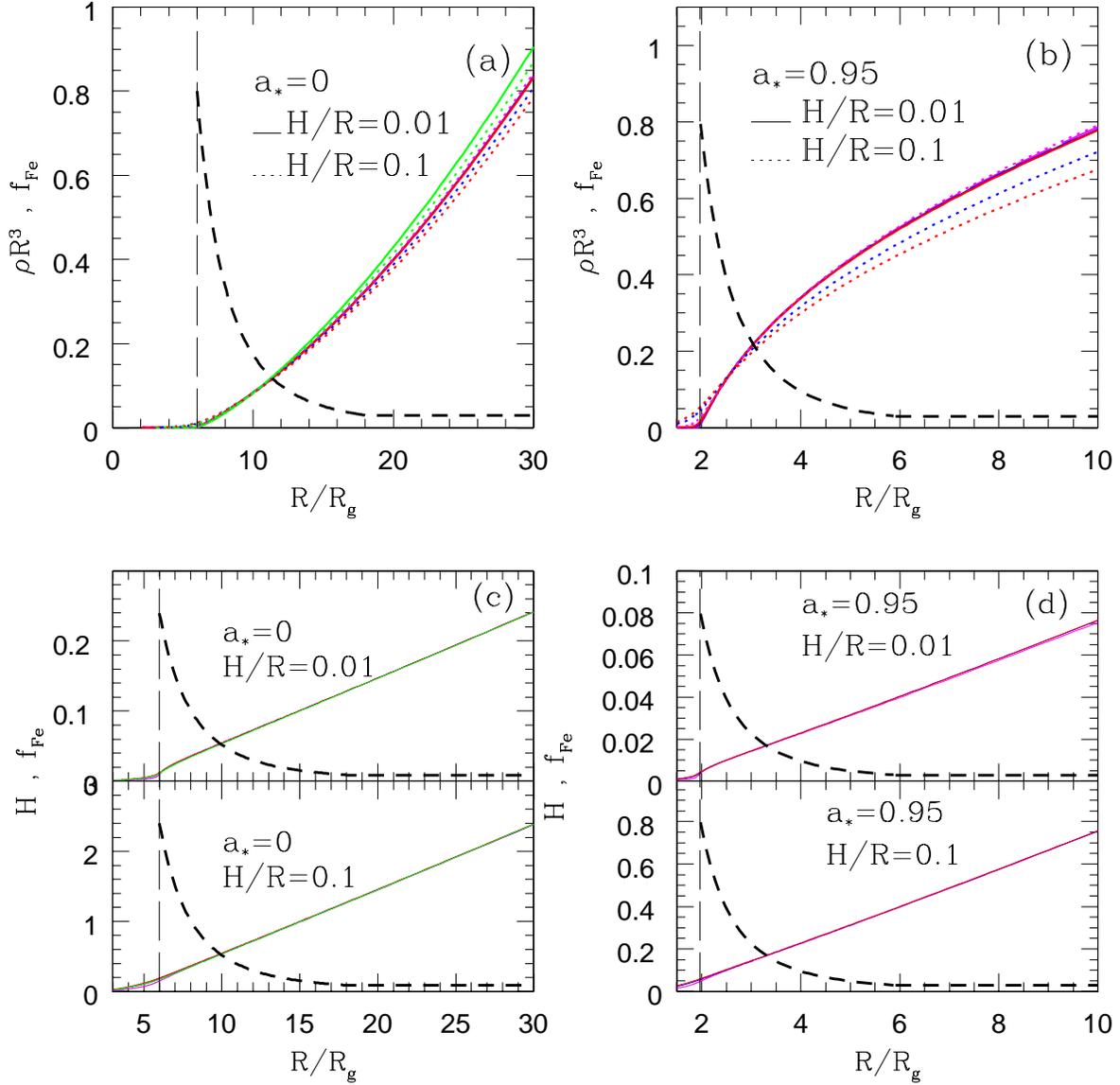} \caption{Profiles of $\rho(R)R^3$
  (panels $a$ and $b$) and disk thickness $H$ (panels $c$ and $d$) in
  the inner regions. For the line types/colors defining the various
  models and the location of the ISCO, see Figures 1 and 2.
  Superimposed on our models in all of the panels is a thick black
  line that schematically represents the emissivity profile
  efficiency, $f_{\rm Fe}$, assumed in the iron line work in which the
  emissivity cuts off abruptly inside the ISCO and falls off as a
  steep power law outside the ISCO (eq. 23).  Panel $a$ shows $\rho(R)
  R^3$ as a function of radius for a non-spinning BH.  Panel $b$ shows
  a similar plot for $a_*=0.95$.  Panels $c$ and $d$ show disk
  thickness $H$ as a function of radius for $a_*=0$ and 0.95, for
  disks with asymptotic values of $H/R = 0.01$ and 0.1. In panels $a$
  and $b$, the normalizations used for $H/R=0.01$ and $\HR=0.1$ are
  different.}
\end{figure}

\newpage
\begin{figure}
  \figurenum{4} \plotone{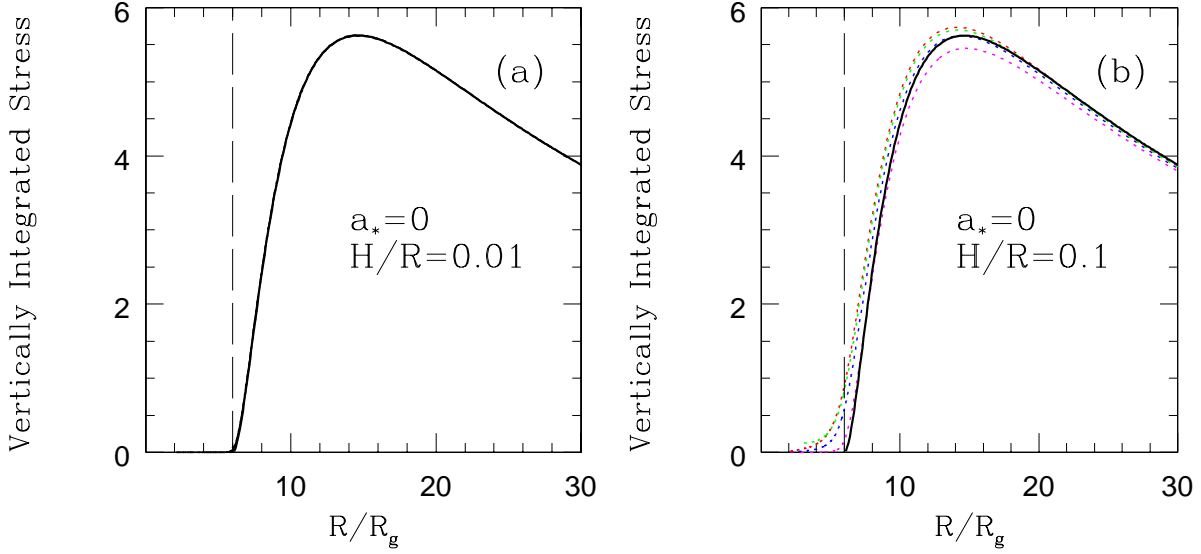} \caption{Vertically integrated
    stress $2 H \alpha P (\times 10^4)$ for a non-spinning BH.  In
    both panels the standard disk model is plotted as a thick solid
    line.  For the line types/colors defining the various models and
    the location of the ISCO, see Figure 1.  ($a$) For $H/R = 0.01$,
    all four models are seen as indistinguishable from the standard
    model.  ($b$) For the thicker disk, the models can be cleanly
    distinguished inside $R \sim 15R_{\rm g}$.  All numerical values
    correspond to $G=c=M=1$, $\dot{M}=1$.}
\end{figure}

\newpage
\begin{figure}
  \figurenum{5} \plotone{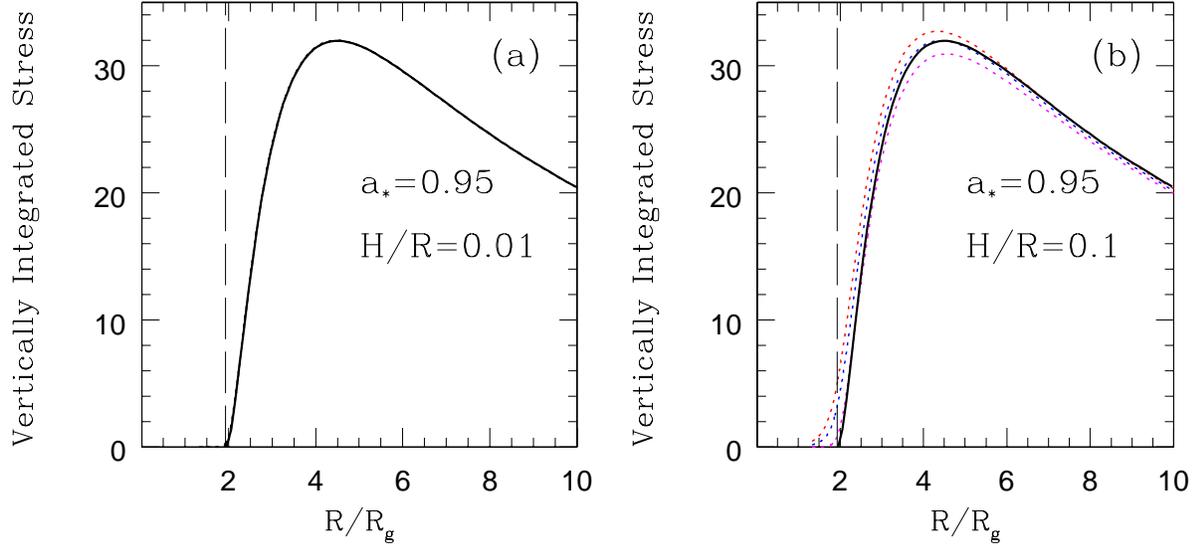} \caption{Similar to Figure 4, but
    for $a_*$=0.95.  As in Figure 2, there is no variable-\al~model, and
    the ISCO is located at $R=1.937 R_{\rm g}$.}
\end{figure}

\newpage
\begin{figure}
  \figurenum{6} \plotone{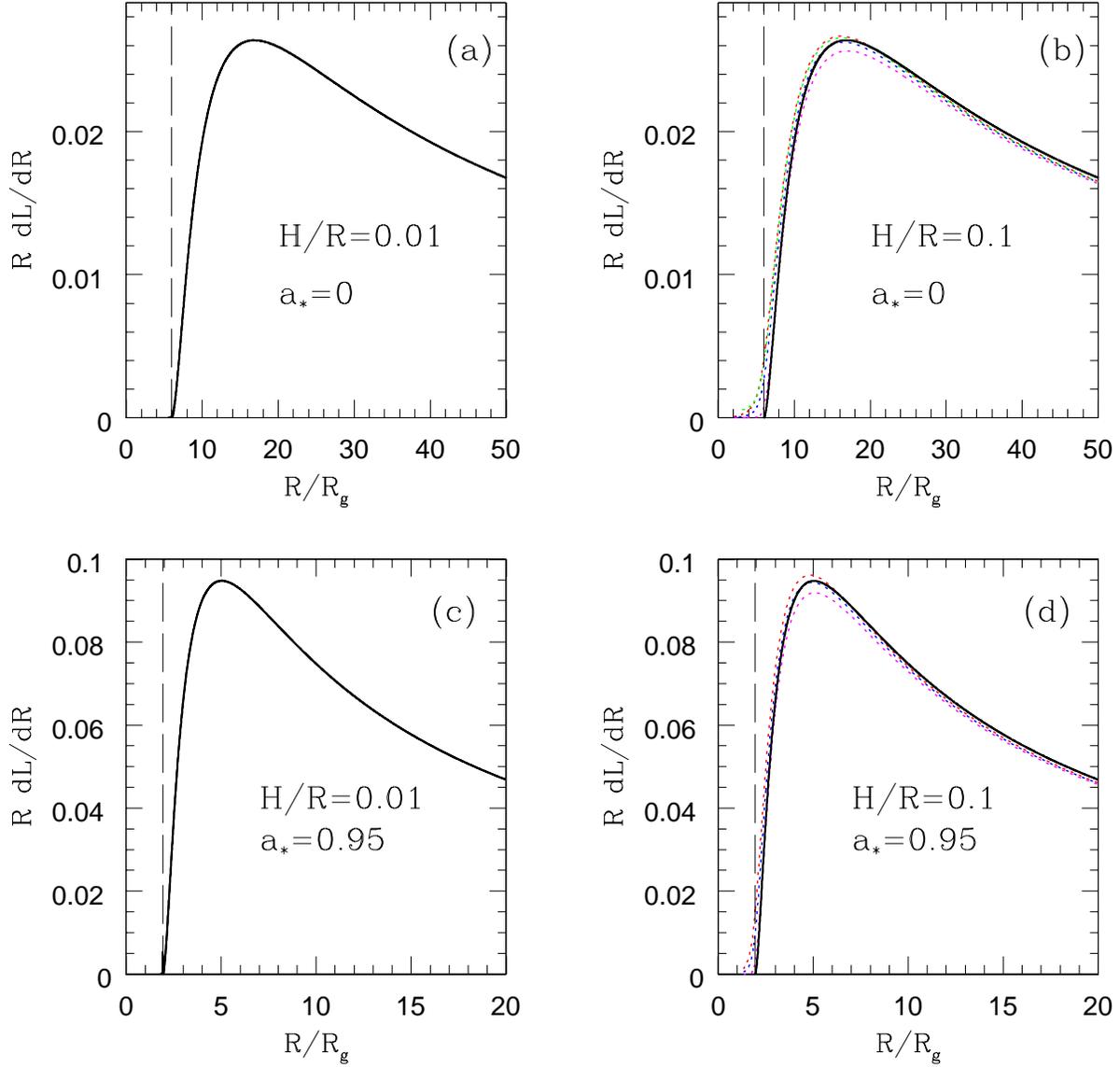} \caption{Rate of energy dissipation
    $R dL/dR$ as a function of radius $R$ for models of a non-spinning
    BH ($a$ and $b$) and a spinning BH ($c$ and $d$). For the line
    types/colors defining the various models and the location of the
    ISCO, see Figures 1 and 2.  The models shown for the thinner disks
    coincide with the standard disk model, whereas the thicker disks
    deviate somewhat from the standard model.}
\end{figure}

\newpage
\begin{figure}
  \figurenum{7} \plotone{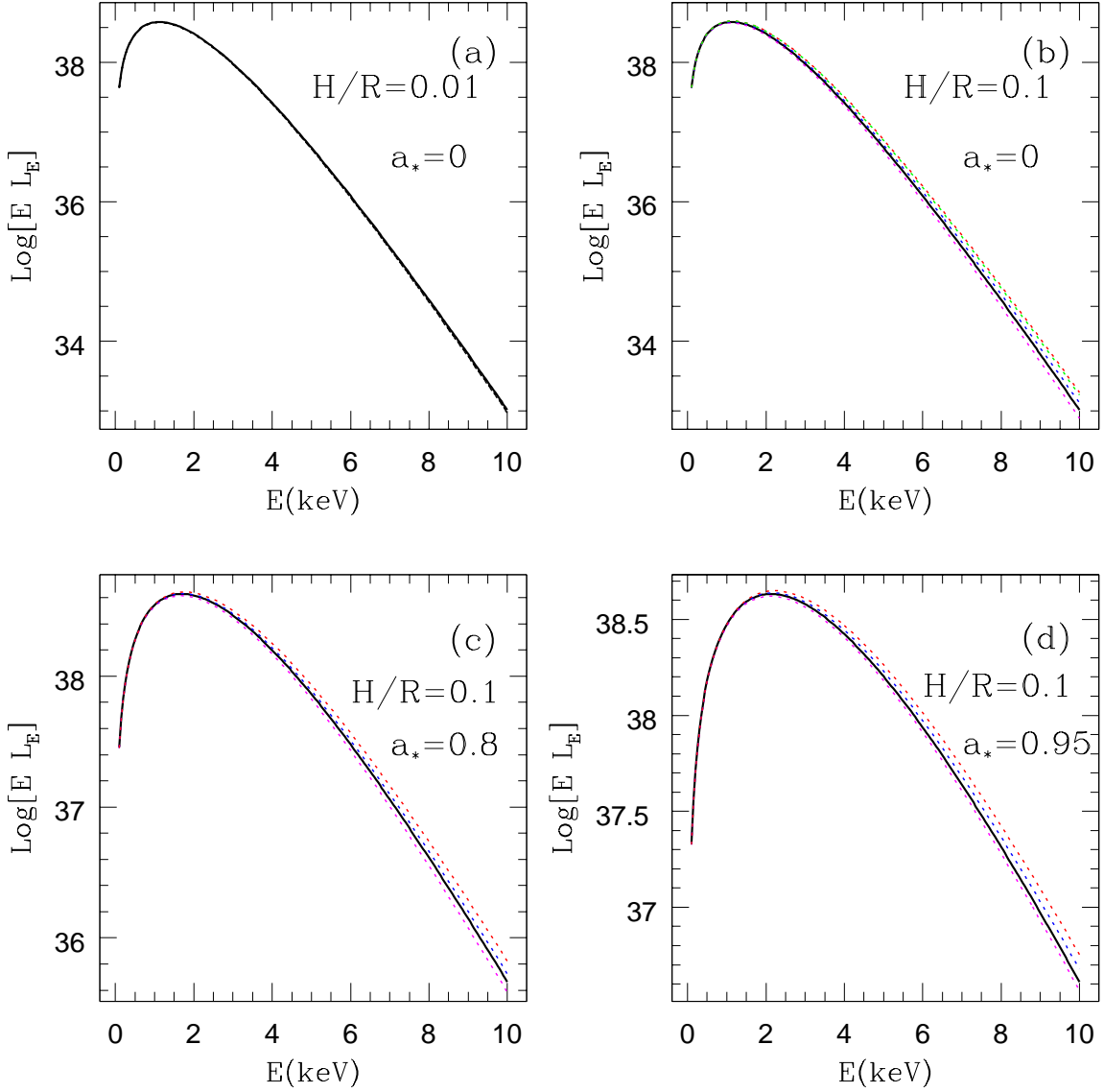} \caption{Spectra corresponding to
    the numerical disk models described in this paper for models of a
    non-spinning BH ($a$ and $b$) and spinning BHs with thicker disks
    for $a_* = 0.8$ ($c$) and $a_* = 0.95$ ($d$).  For the line
    types/colors defining the various models, see Figures 1 and 2.
    Again, the different models are essentially indistinguishable in
    the case of the thin disk ($a$).}
\end{figure}

\newpage
\begin{figure}
  \figurenum{8} \plotone{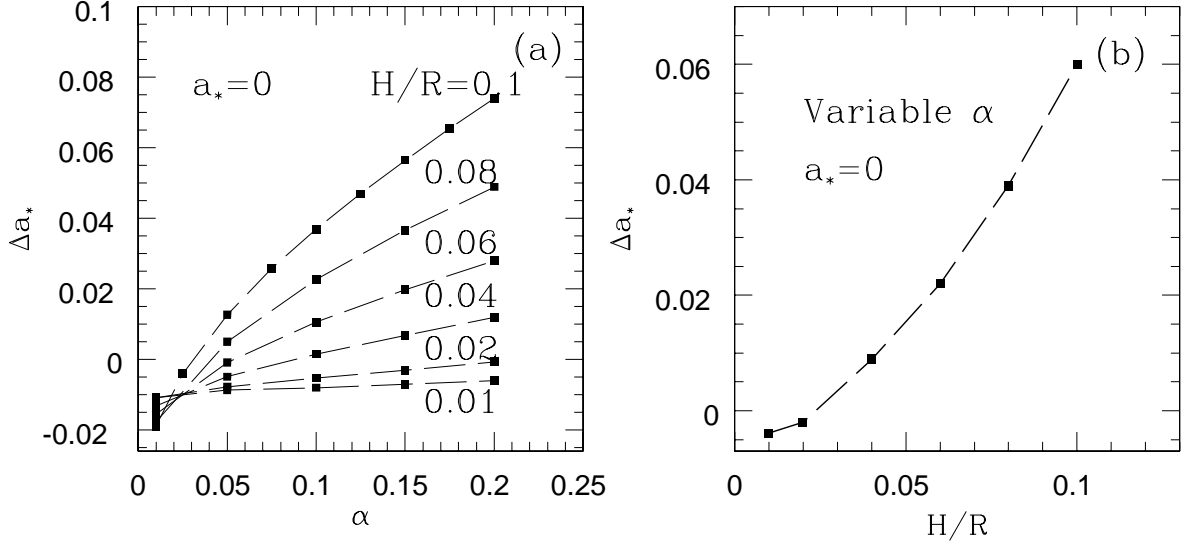} \caption{($a$) Error in the spin
    estimate as a function of $\alpha$ for a BH with a true spin
    parameter of $a_{*} =0$.  The quantity $\Delta a_*$ is equal to
    the value of $a_*$ obtained from fitting the model spectrum minus
    the true $a_{*}$.  The different curves correspond to different
    relative thicknesses $H/R$ of the disk.  ($b$) Error in the spin
    estimate for the variable-$\alpha$ profile as a function of disk
    thickness.  In panel a, note the variable offset from zero error
    that occurs near \al~= 0.01, which is not visible in the previous
    figures for which the the different \al\ models nearly coincide
    with the standard disk model.}
\end{figure}

\newpage
\begin{figure}
  \figurenum{9} \plotone{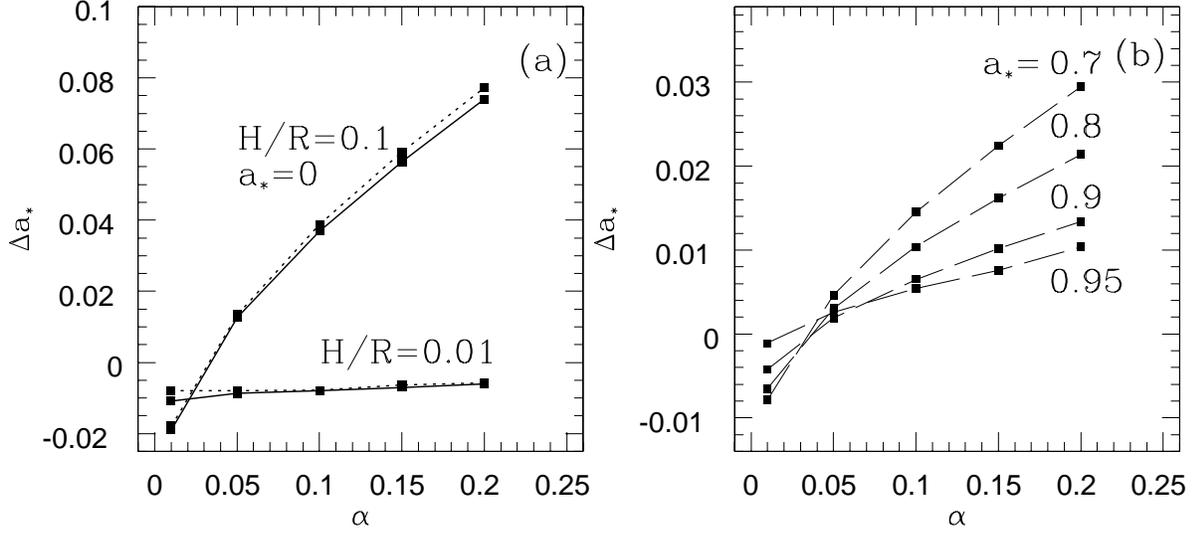} \caption{($a$) Error in the spin
    estimate $\Delta a_*$ for a non-spinning BH, calculated using
    spectral fitting and eq.~(27) (solid lines) and from eq.~(29)
    (dotted lines).  The agreement is very good, showing that the
    simpler approach via eq.~(29) is quite accurate.  ($b$) $\Delta a_*$
    as a function of $\alpha$ for BHs with different values of $a_*$,
    calculated using equation~(29).}
\end{figure}

\newpage
\begin{figure}
  \figurenum{10} \plotone{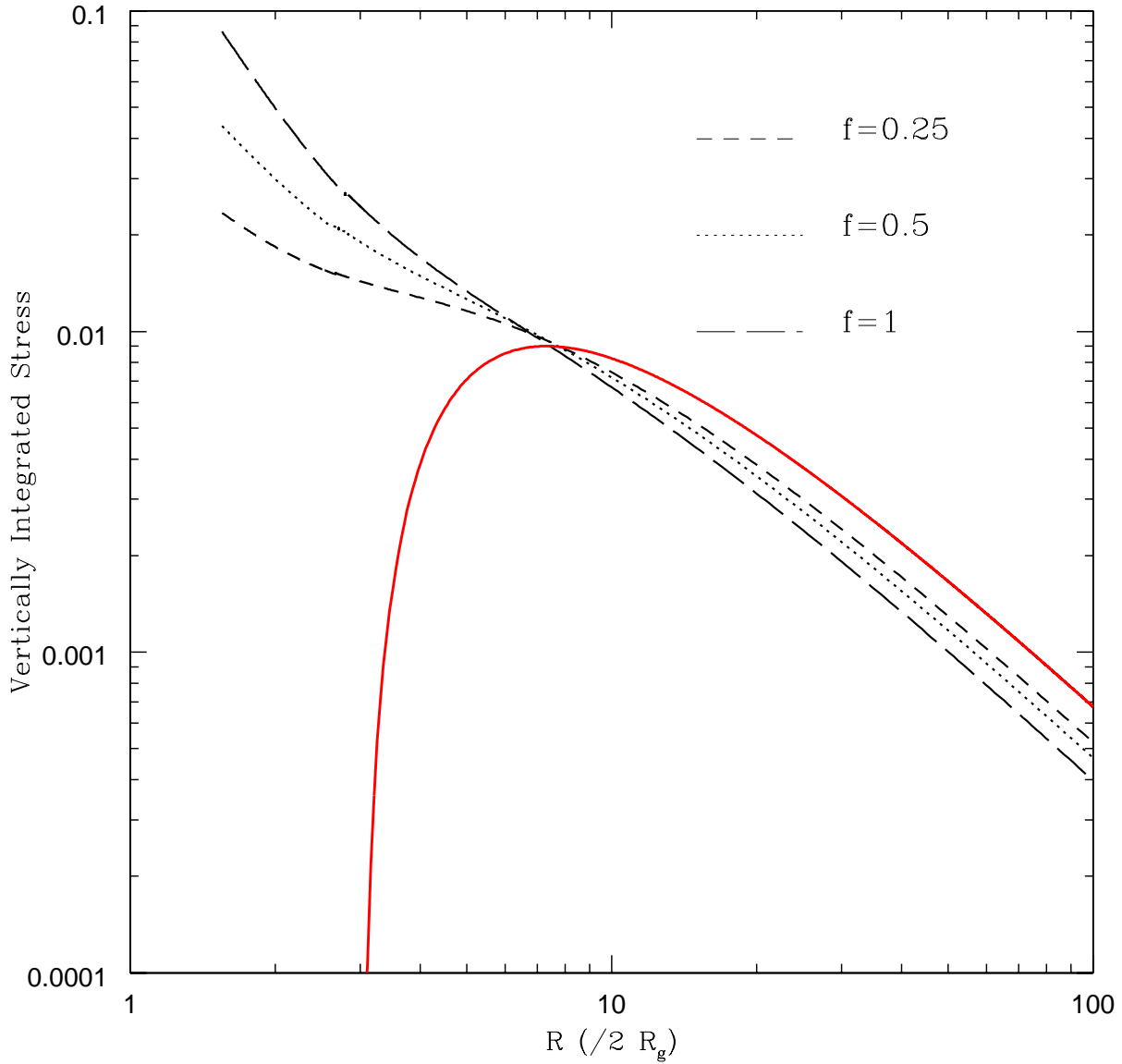} \caption{The thick red solid line
    shows the vertically integrated stress profile as predicted by the
    standard disk model for a non-spinning BH (PW80 potential).  The
    other lines show the stress profiles of three hydrodynamic disk
    models with advection parameter values $f=0.25$ (short-dashed line),
    $f=0.5$ (dotted line) and $f=1$ (long-dashed line).  All three models
    use the variable $\alpha$ prescription (eq. 22), and their stress
    profiles have been scaled to match the standard model at
    $R=15R_g$.  A logarithmic scale has been used to facilitate
    comparison with the MHD simulation result shown in Fig. 10 of
    HK02.}
\end{figure}

\end{document}